\documentstyle[11pt,twoside,colloqOHP2010,epsfig]{article}
\def\mjup{ M_{\rm J}}

\def\mearth{\,{ M}_\oplus}

\def\beq{\begin{equation}}
\def\eeq{\end{equation}}
\def\simgr{\,\hbox{\hbox{$ > $}\kern -0.8em \lower 1.0ex\hbox{$\sim$}}\,}
\def\simle{\,\hbox{\hbox{$ < $}\kern -0.8em \lower 1.0ex\hbox{$\sim$}}\,}

\def\Rp{R_{\mathrm{p}}}
\def\Mp{M_{\mathrm{p}}}

\def\Ms{M_\star}

\def\es{\varepsilon_\star}

\begin{document}
\title{The radius anomaly in the planet/brown dwarf overlapping mass regime.}
\author{J\'er\'emy Leconte$^{1}$,  Gilles Chabrier$^{1}$, Isabelle Baraffe$^{1,2}$ \&  Benjamin Levrard$^{1}$
}

\affil{$^1$  \'Ecole Normale Sup\'erieure de Lyon, 46 all\'ee d'Italie, F-69364 Lyon cedex 07, France; \\ 
 Universit\'e Lyon 1, Villeurbanne, F-69622, France; CNRS, UMR 5574, Centre de Recherche Astrophysique de Lyon;\\
 (jeremy.leconte, chabrier, ibaraffe@ens-lyon.fr)}
\affil{$^2$
School of Physics, University of Exeter, Stocker Road, Exeter EX4 4PE, UK} 
\begin{abstract}
The recent detection of the transit of very massive substellar companions (Deleuil et al. 2008; Bouchy et al. 2010; Anderson et al. 2010; Bakos et al. 2010)
 provides a strong constraint to planet and brown dwarf formation and migration mechanisms. Whether these objects are brown dwarfs originating from the gravitational collapse of a dense molecular cloud that, at the same time, gave birth to the more massive stellar companion, or whether they are planets that formed through core accretion of solids in the protoplanetary disk can not always been determined unambiguously and the mechanisms responsible for their short orbital distances are not yet fully understood.
In this contribution,  we examine the possibility to constrain the nature of a massive substellar object from the various observables provided by the combination of Radial Velocity and Photometry measurements (e.g.~$\Mp,~\Rp,~\Ms,~\mathrm{Age},~a,~e...$).

In a second part, developments in the modeling of tidal evolution at high eccentricity and inclination - as measured for HD 80\,606 with $e=0.9337$ (Naef et al. 2001)
, XO-3 with a stellar obliquity $\es>37.3\pm3.7$ deg (H\'ebrard et al. 2008; Winn et al. 2009)
 and several other exoplanets - are discussed along with their implication in the understanding of the radius anomaly problem of extrasolar giant planets.

\end{abstract}

\section{Quantifying the radius anomaly}
\label{sec:quant}

\begin{figure}[htbp] 
 \centering
 \resizebox{.8\hsize}{!}{\includegraphics{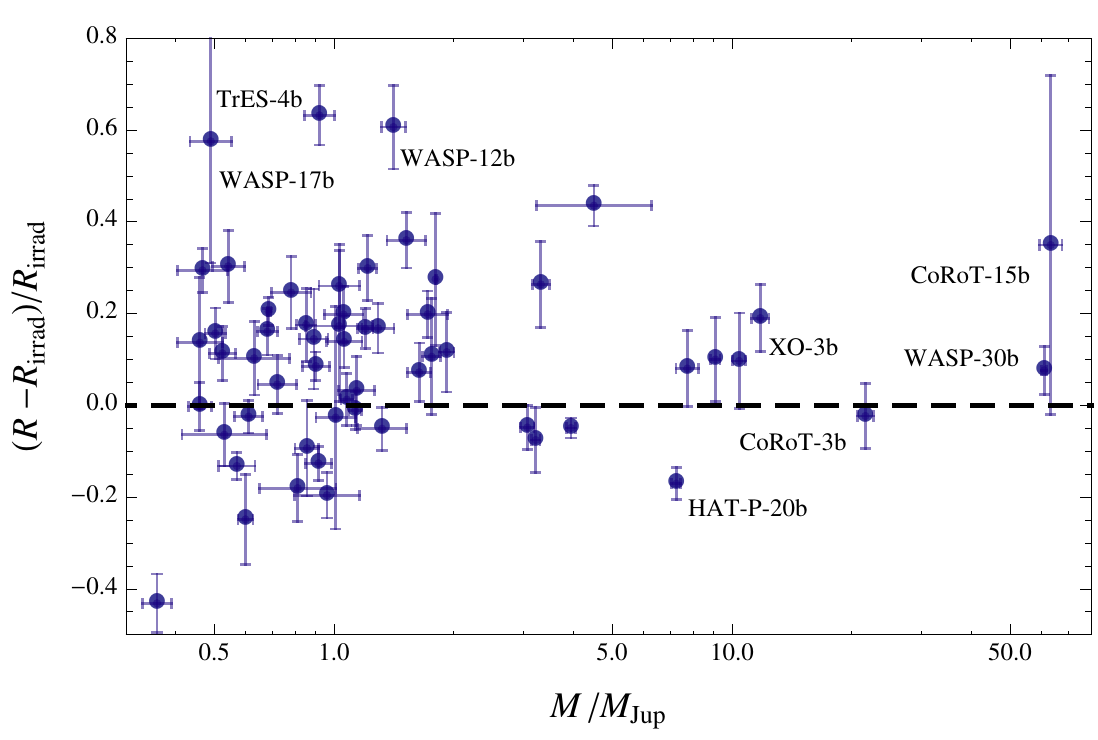}}
  \vspace{-10pt}
 \caption{Relative radius excess between the observationally and the theoretically determined values for 57 transiting systems. Objects significantly above the dashed line are considered to be anomalously bloated compared with the prediction of the regular evolution of an irradiated gaseous planet. All the objects below this line can be explained by a heavy material enrichment in the planet's interior (Baraffe et al. 2008; Leconte et al. 2010).
 }
   \vspace{-10pt}
 \label{fig:RsurRirrad}
\end{figure}

Because the radius of a gaseous giant planet is not set only by its mass, but strongly depends also on the object's composition, age and irradiation history, the mass-radius diagram only gives a limited view of the constraints offered by the observation of transiting systems.
In order to quantify the \textit{radius anomaly} of many "Hot Jupiters" and study the possibility of such an anomaly for the more massive objects, we computed the radius predicted by our standard model ($R_{\mathrm{irrad}}$), described in Baraffe, Chabrier \& Barman (2008) and Leconte et al. (2009),
for detected transiting planets with $\Mp>0.3\mjup$ (about a Saturn mass). Results are summarized in Fig.\,\ref{fig:RsurRirrad}.

\subsection{Inflated planets}\label{sec:inflated}

The objects significantly above the $R=R_{\mathrm{irrad}}$ line show suggest that a missing physical mechanism which is either injecting energy in the deep convective zone or reducing the net outward thermal flux is taking place in these objects. 
Several possibilities 
have been suggested  to explain this mechanism. Tidal heating due to circularization of the orbit, as originally suggested by Bodenheimer, Lin \& Mardling (2001)
is discussed in more detail in \S\ref{sec:tides} Other proposed mechanisms include downward transport of kinetic energy originating from strong winds generated at the planet's surface (Showman \& Guillot 2002)
, enhanced opacity sources in hot-Jupiter atmospheres (Burrows et al. 2007)
, ohmic dissipation in the ionized atmosphere (Batygin \& Stevenson 2010)
or (inefficient) layered or
oscillatory convection in the planet's interior (Chabrier \& Baraffe 2007).

\subsection{The Brown Dwarf/Planet overlapping mass regime}
\label{sec:BDGP}

The recent transit detection of massive companions in the substellar regime  ($ 5\, \mjup \simle\Mp \simle M_{\mathrm{HBMM}}=0.075\,M_\odot$), where $M_{\mathrm{MMHB}}$ is the hydrogen burning minimum mass, in close orbit to a central star
raises the questions about their very nature: planet or brown dwarf ? 


The brown dwarf status of objects such as CoRoT-15\,b (Bouchy et al. 2010)
and WASP-30\,b (Anderson et al. 2010)
can not be questioned given their mass close to $M_{\mathrm{MMHB}}$. Such masses can not be produced by the core accretion mechanism for planet formation. Interestingly enough, these objects are not significantly inflated at the 1-2 $\sigma$ level (because of the large uncertainty in the age and radius determination, especially for CoRoT-15\,b).

\begin{figure}[htbp] 
 \centering
 \resizebox{.6\hsize}{!}{\includegraphics{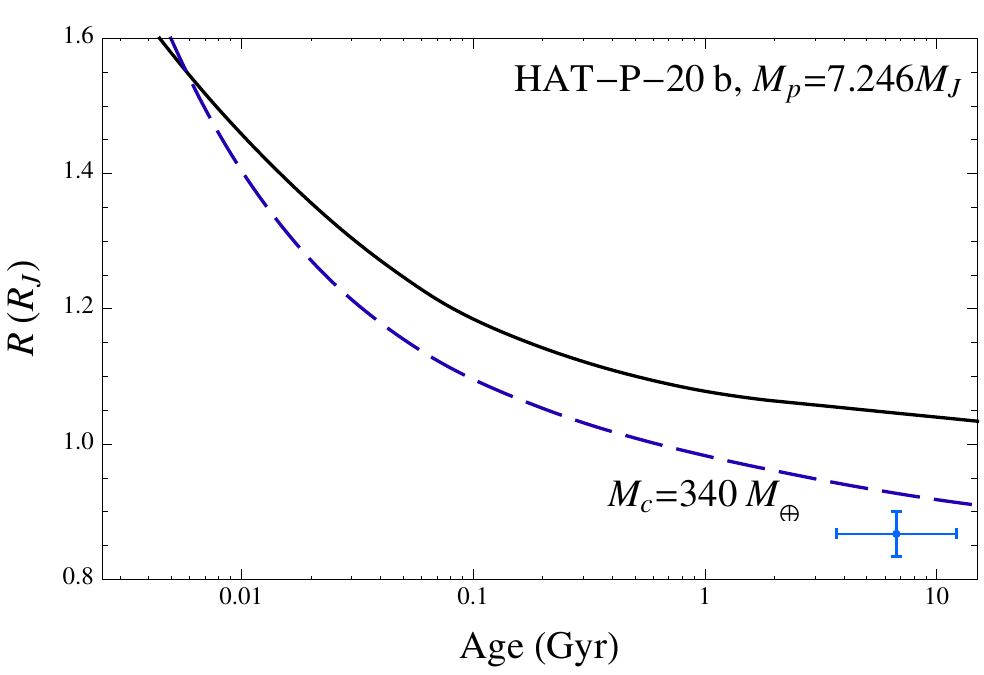}}
  \vspace{-5pt}
 \caption{Radius evolution of a 7.2$\mjup$ planet model with solar composition (Solid) and with a 340$\mearth$ water core (Dashed) compared with the 1$\sigma$ error bars}
 \label{fig:hatp20}
  \vspace{-7pt}
\end{figure}

For objects such as CoRoT-3\,b (Deleuil et al. 2008),
HAT-P-20\,b (Bakos et al. 2010)
and planets in the $ 5 \mjup \simle\Mp \simle 20 \mjup$ range, the situation is more ambiguous.
Studies of low mass 
stars and brown dwarfs in young clusters suggest a continuous mass function down to 
$\sim 6\,\mjup$ (Caballero et al. 2007),
indicating that the same formation process
responsible for star formation can produce objects down to a few Jupiter masses, as supported by analytical theories (Padoan \& Nordlund 2004; Hennebelle \& Chabrier 2008).
As discussed in Leconte et al. (2009),
this ambiguity can be resolved in the favorable case where the observed radius is significantly {\it smaller} than predicted for solar or nearly-solar metallicity (irradiated) objects. This indeed reveals the presence of a significant global amount of heavy material in the transiting object's interior (Baraffe et al. 2006; Fortney, Marley \& Barnes 2007; Burrows et al. 2007; Baraffe et al. 2008; Leconte et al. 2009; Baraffe, Chabrier \& Barman 2010),
a major argument in favor of the core-accretion planet formation scenario. Thus, if the status of CoRoT-3\,b cannot be yet determined, HAT-P-20\,b (if confirmed) shows evidences of a planetary nature. Estimating a rough upper limit on the amount of heavy elements available in the disc to form planets,
\begin{equation}
M_Z\approx \eta  \cdot Z \cdot (f \cdot M_\star),
\end{equation}
with $f\cdot$$M_\star$ the maximum mass for a stable disk ($\simle0.1 M_\star$), $Z$ the metal mass fraction of the star and $\eta\approx 30\%$ the accretion efficiency rate, yields $M_Z\approx 340 \mearth$ for HAT-P-20\,b. As shown on Fig.\,\ref{fig:hatp20}, such a large enrichment in heavy elements is required in the modeling of the object to match the observed radius at the 1-1.5$\sigma$ level. The small remaining discrepancy can be reduced by considering an enrichment in silicate, iron or a mixture of those elements instead of pure water (Baraffe et al. 2008).
While massive objects like HAT-P-20\,b are at the upper limit of the mass distribution predicted by the core accretion scenario, their large metal enrichment certainly excludes formation by gravitational collapse.
 
On the opposite, if a physical mechanism is missing in current planet cooling models, observed radii {\it larger} than predicted by the models do not necessarily imply a lack or a small amount of heavy material. For such cases, the nature of the object remains ambiguous, if only based on the determination of its mean density. 


\section{Uncertainties in tidal theory}\label{sec:tides}

Tidal heating has been suggested by several authors to explain the anomalously large radius of some giant close-in observed exoplanets (Bodenheimer et al. 2001, Jackson, Greenberg \& Barnes 2008; Miller, Fortney \& Jackson 2009; Ibgui, Spiegel \& Burrows 2009).
Their best case scenario consists in a planet left on a wide, very eccentric orbit by an early event during its formation, whose orbit is slowly decaying due to tidal dissipation, leading to a circularization on a timescale of a few Gyr's. This slow circularization is due to the use of a tidal model based on a {\it quasi circular approximation} and thus truncated at $2^{\mathrm{nd}}$ order in eccentricity.
 
This quasi circular approximation, developed to study the tidal evolution of the {\it solar system} planets (Goldreich \& Soter 1966; Ferraz-Mello, Rodr\'{\i}guez \& Hussmann 2008),
which have very low eccentricities, is valid only in this limit. 
In the context of exoplanetary systems where (today) high eccentricities are common and initial high eccentricities are very likely, as inferred from non-transiting planets observed by radial velocity, this quasi circular approximation is no longer correct, as demonstrated in Wisdom (2008)
and Leconte et al. (2010)
and summarized below. 

\begin{figure}[htbp] 
 \centering
 \resizebox{.7\hsize}{!}{\includegraphics{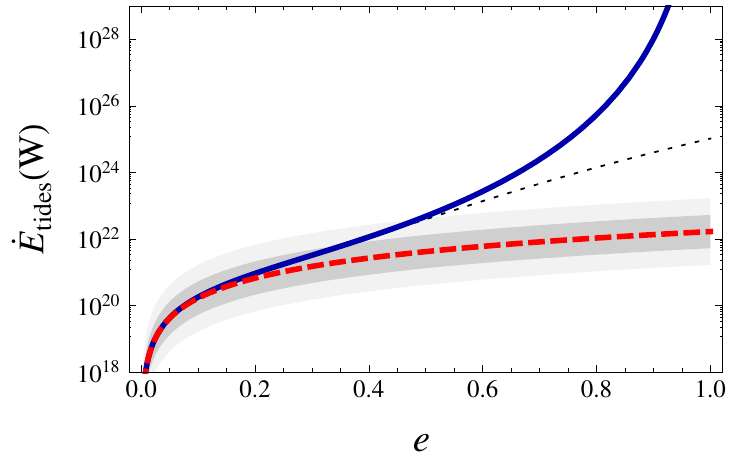}}
 \vspace{-10pt}
 \caption{Tidal energy dissipation rate in a pseudo-synchronized planet as a function of the eccentricity calculated with the complete formula (\textit{curve}), with the $e^2$-truncated formula (\textit{dash}) and to $e^{10}$ (\textit{Dotted}). The ratio of the two curves only depends on the eccentricity and not on the system's parameters. The inner (outer) shaded area shows the uncertainty in the heating when allowing the dissipation parameter to vary within one (two) order of magnitude. The actual values were derived using HD 209\,458\,b parameters.}
 \label{fig:edot}
 \vspace{-10pt}
\end{figure}

Present analytical theories for tidal interaction are all based on the equilibrium tides and weak friction approximation, since no adequate theory for dynamical tides presently exists. These theories differ in two ways
\begin{itemize}
\item (i) their parametrization of the dissipative processes. The most common prescriptions are either a constant phase lag (constant-$Q$) model or a constant viscosity or time lag (constant-$\Delta t$) model.
\item (ii) their mathematical treatment of the geometry of the orbits: perturbative developments around the coplanar/circular keplerian orbits or closed formulae, valid for any eccentricity.
\end{itemize}
While these two sources of differences between the models are completely different in nature, they are often, erroneously, mixed together. Indeed, only the constant time lag model, because of the linear dependence of the phase lag upon the time lag in this model, allows the calculations to be carried out in terms of closed formulae for any eccentricity. High order calculations in eccentricity in the framework of the constant-$Q$ model are very cumbersome
(see Ferraz-Mello et al. 2008).

 As demonstrated by Wisdom (2008)
and Leconte et al. (2010), 
even if large uncertainties remain on the quantification of the dissipative processes, the discrepancies arising from the differences in the treatment of the orbital geometry at moderate to high eccentricities ($e\simgr 0.2$-0.3) can become dominant by orders of magnitude. This is summarized on Fig.\,\ref{fig:edot} which compares the tidal heating given by the constant time lag model of Leconte et al. (2010)
(solid curve) and by the quasi circular approximation of Peale \& Cassen (1978)
(dashed curve). In comparison, the inner (outer) shaded area illustrates the impact of the uncertainty in the heating when allowing the tidal dissipation parameter to vary by one (two) order of magnitude. For $e>0.4$, we see that high order terms in $e$ yield a contribution which is larger than the uncertainty in the quantification of the dissipative processes. Such a behavior at high eccentricity is well understood in the context of celestial mechanics and is due to the slow convergence of elliptical expansion series (Danjon 1980; Cottereau, Aleshkina \& Souchay 2010).

  Therefore, calculations based on constant-$Q$ models truncated at the order $e^2$ cannot be applied to (initial or actual) eccentric orbits larger than about 0.2-0.3, a common situation among detected exoplanetary systems. This implies a major caveat in previous calculations coupling thermal and orbital evolutions. In particular, as illustrated in Fig. 3, using a $e^2$-truncated model leads to a severely underestimated tidal dissipation timescale at large eccentricity, and thus to an overestimated amount of dissipated tidal energy in exoplanet interiors at present ages (Leconte et al. 2010). 
Revisiting the viability of the tidal heating hypothesis to explain the anomalously large Hot Jupiter radii with the Hut complete tidal model, Leconte et al. (2010)
 (see also Hansen 2010)
 showed that, although tidal friction indeed provides a possible explanation for some transiting systems, the tidal heating hypothesis fails to explain the radii of extremely bloated planets like - among others - HD 209\,458\,b, TrES-4\,b, WASP-4\,b or WASP-12\,b. The main reason is the early circularization of the orbit of these systems which are thus insufficiently heated at a late epoch.
Note that we have only considered a two body problem. The presence of a third body able to excite eccentricity in a massive giant planet for several gigayears would provide an other explanation. Accurate observations are necessary to support or exclude the presence of such undetected close low-mass or distant massive companions.

\acknowledgments{This work was
supported by the Constellation european network MRTN-CT-2006-035890, the french ANR "Magnetic Protostars and Planets"
(MAPP) project and the "Programme National de Plan\'etologie" (PNP) of CNRS/INSU. We acknowledge the use of the \textit{www.exoplanet.eu} database. J.L. wishes to thank L. Cottereau for insightful discussions concerning current developments in Celestial Mechanics.}

\end{document}